# EXPERIMENTAL DESIGNS FOR MULTIPLE-LEVEL RESPONSES, WITH APPLICATION TO A LARGE-SCALE EDUCATIONAL INTERVENTION[1]

By Brenda Jenney and Sharon Lohr

*Arizona State University*

Educational research often studies subjects that are in naturally clustered groups of classrooms or schools. When designing a randomized experiment to evaluate an intervention directed at teachers, but with effects on teachers and their students, the power or anticipated variance for the treatment effect needs to be examined at both levels. If the treatment is applied to clusters, power is usually reduced. At the same time, a cluster design decreases the probability of contamination, and contamination can also reduce power to detect a treatment effect. Designs that are optimal at one level may be inefficient for estimating the treatment effect at another level. In this paper we study the efficiency of three designs and their ability to detect a treatment effect: randomize schools to treatment, randomize teachers within schools to treatment, and completely randomize teachers to treatment. The three designs are compared for both the teacher and student level within the mixed model framework, and a simulation study is conducted to compare expected treatment variances for the three designs with various levels of correlation within and between clusters. We present a computer program that study designers can use to explore the anticipated variances of treatment effects under proposed experimental designs and settings.

**1. Introduction.** Randomized experiments are frequently recommended for evaluating educational studies [Cook and Payne (2002)]. The What Works Clearinghouse (2006, page 5), guidelines state: "Studies that *Meet Evidence Standards* are well-designed and implemented randomized controlled trials." Boruch (2002), page 38, comments that although randomized trials have "been slow to come to the field of education," they provide the best way of ascertaining which interventions are truly effective

Received December 2007; revised October 2008.
[1]Supported in part by NSF Grants EHR-0412537 and SES-0604373.

for helping students. Randomized studies can provide evidence of a causal relationship between an intervention and results, and they are frequently cited by educational reformers [Gueron (2005)]. There are many ways, however, of conducting a randomized trial, and it is desired to have a design that provides as much information as possible using the available resources. In this paper we study the efficiency of randomized designs for a situation in which teachers at multiple schools are randomized to treatments, and the impact of the educational treatment can be measured at multiple levels: in our case, at both the teacher and the student level.

The research in this paper was motivated by the problem of evaluating the effects of Project Pathways [CRESMET, Arizona State University (2007)]. Project Pathways is a professional development program for secondary STEM (Science, Technology, Engineering and Mathematics) teachers, with the immediate goal of increasing teachers' conceptual and pedagogical knowledge of STEM topics. The primary intervention of Project Pathways is a set of four courses taken by secondary STEM teachers in schools located in Maricopa County, Arizona. The courses are developed around the themes of functions and proportional reasoning. Course 1 concentrates on the mathematics, and courses 2–4 integrate the mathematical concepts with biology, physics, chemistry, geology and engineering. It is hypothesized that the teachers' increased understanding will lead to increased knowledge and achievement for the students who take classes from those teachers. Thus, while the treatment is administered to teachers, effects of the treatment need to be evaluated at both the teacher level and the student level. Students will likely take classes from several teachers over several semesters, so the data structure will not be completely hierarchical. In a completely hierarchical structure the students would be nested in teachers. An additional complication is that Project Pathways has no input on assigning students to teachers, so the design needs to be robust to possible self-selection by students. This research aims to provide some guidance for choice of randomized experimental design in multi-school studies such as Project Pathways when the effect of the intervention can be measured at more than one level. We develop theoretical results comparing efficiencies of designs in a general setting. We also present a computer program, *multleveldesign*, that designers can use to plan a study tailored to their circumstances.

While there have been many small-scale randomized studies performed within schools, and many randomized studies relating to education programs that focus on students' physical and mental health, as well as tobacco, drug and alcohol use prevention programs, there have been relatively few whole school reforms tested with randomized studies [Cook (2003)]. As more emphasis is placed on rigorous evaluations in education, the large-scale trials that are common in medical research should become more prevalent when studying educational innovations. One program currently under evaluation by randomized study is the reading program Success for All, which currently tracks the progress of students in 41 schools [Borman et al. (2005)]. Randomized studies have been found to give more reliable results in other fields, such as psychiatry [Johnson (1998)] and criminal justice [Berk et al. (2003)]. Cook (2005) provides a synthesis of the most commonly encountered problems in cluster randomized designs, as well as the merits of cluster-based experiments in the social sciences. Reasons to consider cluster randomization are also explored by Gail et al. (1996). For a discussion of optimal design when budget constraints are present, see Moerbeek, van Breukelen and Berger (2000) and Raudenbush and Liu (2000).

A number of researchers have studied the merits of different randomized designs in the hierarchical setting when the response is measured at one level. Raudenbush (1997) gives guidance on how to optimally design cluster-randomized studies, based on a criterion of minimizing the standard error of the treatment contrast. Bloom, Bos and Lee (1999) study the power of cluster-randomized designs and note that, given the same number of individuals in a program, these types of designs produce a smaller effective sample size. Bloom, Bos and Lee (1999), Raudenbush (1997) and Bloom, Richburg-Hayes and Black (2005) advocate the use of covariates that are related to the measured response to increase the statistical power in cluster-randomized designs. Moerbeek, van Breukelen and Berger (2000) derive the relative efficiency of randomize-by-school vs. randomize-teacher-within-school designs for evaluating the impact of treatment on teachers. Moerbeek (2005) studies the effect of contamination—when some teachers in the control group adopt the intervention method—on power. The efficiency of designs for multiple response levels when the data structure is not completely hierarchical, however, has not been previously explored.

To provide guidance on study design for multilevel-response studies of this type, we look at the efficiencies for measuring program impact on teachers and students for three randomization schemes for the intervention: (1) randomly assigning schools, with all their teachers and students, to the experimental or control groups, (2) randomly assigning half of the teachers within each school to the experimental group (a randomized block design for teachers), and (3) randomly assigning teachers regardless of school to the two groups (a completely randomized design for teachers). In Section 2 we construct a unified mixed model framework for responses of teachers and students, and in Section 3 we study the relative efficiencies of the three designs for assessing treatment effects at teacher and student levels. In Section 4 we explore effects of possible treatment contamination, noncompliance or attrition on the relative efficiencies. Section 5 describes a computer program that may be used to simulate the distribution of the anticipated variance of the various designs for measuring the impact of the intervention on both teachers and students. In Section 6 we discuss implications of the results for evaluation design choice.

In this paper we express all results in terms of educational experiments for ease of interpretation, but the research results apply to many other settings as well. A multicenter clinical trial involving a new physical therapy method may randomize therapists to treatments within centers, or may randomize entire centers with all their staffs to treatments. In this case, it is desired to measure effects of the intervention at both therapist and patient levels. Similarly, a study on most effective police response to domestic violence incidents may randomize treatment assignment at any of several levels: city, police station, police officer or incident. There may be multiple calls to the same household during the study period, so that a household may interact with several police officers. Responses of interest might include officers' knowledge of and actions about domestic violence incidents, subsequent domestic violence reports from a household in the study or incident characteristics. The responses thus occur at three levels of experimental units.

**2. Models for responses.** The goal of the study is to evaluate effects of the intervention on teachers and simultaneously on students in their classes. With that goal in mind, we introduce models that could be used at each level of response. In the following, $\mathbf{I}_k$ is the $k \times k$ identity matrix, $\mathbf{1}_k$ is the $k$-vector of all ones, and $\mathbf{J}_k = \mathbf{1}_k \mathbf{1}_k'$. We assume that there are $a$ schools available for the study, and that school $i$ has $m_i$ teachers and $n_i$ students who could participate.

2.1. *Teacher model.* Let $T_{ij}$ be a response of interest for teacher $j$ at school $i$. $T_{ij}$ might be, for example, the change score on an assessment of content knowledge given before and after the intervention or, alternatively, $T_{ijt}$ could be the score on the assessment at time $t$ in a longitudinal study. If we use a change score as a response, a possible model for the teachers' response is the mixed effects model

$$\mathbf{T}_i = \mathbf{X}_i \boldsymbol{\beta} + \mathbf{1}_{m_i} v_i + \boldsymbol{\varepsilon}_i, \tag{2.1}$$

where $\mathbf{T}_i = (T_{i1}, T_{i2}, \ldots, T_{im_i})'$ is an $m_i$-vector of responses, $\mathbf{X}_i = [\,\mathbf{x}_{i1} \quad \mathbf{x}_{i2} \quad \cdots \quad \mathbf{x}_{im_i}\,]'$ is an $m_i \times p$ matrix of known covariates for the teachers, $\boldsymbol{\beta} = (\beta_1, \beta_2, \ldots, \beta_p)'$ is a $p$-vector of fixed effects, $v_i \sim N(0, \sigma_v^2)$ is a random effect for the school, and $\boldsymbol{\varepsilon}_i \sim N(0, \sigma_\varepsilon^2 \mathbf{I})$ is an $m_i$-vector of random error terms for the teachers. Additionally, $v_i$ and $\varepsilon_{ij}$ are independent for all $i$ and $j$. We assume in this model, then, that teachers from different schools are independent. The last column of $\mathbf{X}_i$ is the treatment assignment for teacher $j$ from school $i$:

$$x_{ijp} = \begin{cases} 1, & \text{if in experimental group,} \\ -1, & \text{if in control group.} \end{cases}$$

The last element of $\boldsymbol{\beta}$, $\beta_p$, is the parameter of interest for assessing the effect of the treatment on the teachers.

Using mixed model theory [Demidenko (2004)], we have, for the setup in (2.1),

$$\mathrm{Cov}(\mathbf{T}_i|\mathbf{X}_i) = \mathbf{V}_i = \sigma_v^2 \mathbf{J}_{m_i} + \sigma_\varepsilon^2 \mathbf{I}_{m_i}.$$

Because the data for all schools are independent of one another, the information for $\beta_p$ is the sum of the information from each school. The information matrix of the generalized least squares estimator $\hat{\boldsymbol{\beta}} = (\sum_{i=1}^a \mathbf{X}_i' \mathbf{V}_i^{-1} \mathbf{X}_i)^{-1} \sum_{i=1}^a \mathbf{X}_i' \mathbf{V}_i^{-1} \mathbf{T}_i$ is

$$(2.2) \qquad \mathcal{I}_T(\hat{\boldsymbol{\beta}}, \mathbf{X}) = \sum_{i=1}^a \mathbf{X}_i' \mathbf{V}_i^{-1} \mathbf{X}_i,$$

where $\mathbf{X} = [\mathbf{X}_1' \cdots \mathbf{X}_a']'$ and

$$(2.3) \qquad \mathbf{V}_i^{-1} = \frac{1}{\sigma_\varepsilon^2} \mathbf{I}_{m_i} - \frac{\sigma_v^2}{\sigma_\varepsilon^2(\sigma_\varepsilon^2 + \sigma_v^2 m_i)} \mathbf{J}_{m_i}.$$

We are primarily interested in the $(p,p)$ entry of the matrix $\mathrm{Cov}(\hat{\boldsymbol{\beta}}) = [\mathcal{I}_T(\hat{\boldsymbol{\beta}}, \mathbf{X})]^{-1}$.

2.2. *Student model.* During the span of the experiment, student $k$ at school $i$ may have classes from one or more of the teachers in the project. Let $Y_{ik}$ be a response measure for student $k$ at school $i$ for $k = 1, \ldots, n_i$. One choice for $Y_{ik}$ might be a change score for an assessment given before and after the intervention.

A model for $Y_{ik}$ needs to allow students to take multiple classes, and to account for dependence among students in the same school and among students who take classes from the same teacher. We propose the following mixed model, related to a model in McCaffrey et al. (2004), for the student's response. The response $Y_{ik}$ depends on characteristics of the student and on characteristics of each teacher who instructs the student $(i, k)$:

$$(2.4) \qquad \mathbf{Y}_i = \mathbf{B}_i \boldsymbol{\gamma} + \mathbf{D}_i (\mathbf{X}_i \boldsymbol{\theta} + \mathbf{t}_i) + \mathbf{1}_{n_i} s_i + \boldsymbol{\eta}_i.$$

Here, $\mathbf{Y}_i = (Y_{i1}, Y_{i2}, \ldots, Y_{in_i})'$ is an $n_i$-vector of student-level responses, $\mathbf{B}_i = [\mathbf{b}_{i1} \mathbf{b}_{i2} \cdots \mathbf{b}_{in_i}]'$ is an $n_i \times q$ matrix of known covariates for students in school $i$, and $\boldsymbol{\gamma} = (\gamma_0, \gamma_1, \ldots, \gamma_{q-1})'$ is a $q$-vector of fixed effects. The $n_i \times m_i$ matrix $\mathbf{D}_i$ describes the assignment of students to teachers: the $(k, j)$ element of $\mathbf{D}_i$ is $d_{ikj}$ = number of classes student $k$ from school $i$ takes with teacher $j$. As in Section 2.1, $\mathbf{X}_i = [\mathbf{x}_{i1} \ \mathbf{x}_{i2} \ \cdots \ \mathbf{x}_{im_i}]$ is an $m_i \times p$ matrix of covariates for teachers at school $i$, whose last column is a vector of treatment indicators. The $p$-vector $\boldsymbol{\theta} = (\theta_1, \theta_2, \ldots, \theta_p)'$ is a vector of fixed effects for the teachers. Since multiple students take classes from teacher $(i, j)$, we include a random effects vector, $\mathbf{t}_i = (t_{i1}, t_{i2}, \ldots, t_{im_i})'$, such that each $t_{ij} \sim N(0, \sigma_t^2)$. We posit an additive model for the effects of teachers on an individual student with element $k$ of $\mathbf{D}_i(\mathbf{X}_i\boldsymbol{\theta} + \mathbf{t}_i)$ representing the additive effect of all of the teachers taken by student $(i, k)$. A student may take classes from any number of the $m_i$ teachers in the school, and may have the same teacher for multiple classes. The model also includes a random effect for the school, $s_i \sim N(0, \sigma_s^2)$, and random error terms for the students, $\boldsymbol{\eta}_i = (\eta_{i1}, \eta_{i2}, \ldots, \eta_{in_i})'$, such that $\eta_{ik} \sim N(0, \sigma_\eta^2)$. Assume that $t_{ij}, s_i$ and $\eta_{ik}$ are mutually independent for all values of the indices $i$, $j$ and $k$.

In practice, one would generally include only teachers participating in the experiment, or teaching classes related to the student outcomes, in the model in (2.4). Nonrelevant teachers—for example, physical education teachers in a study with mathematics outcomes—are likely to have little effect on the response through their physical education classes. If desired, alternative formulations of the $\mathbf{D}_i$ matrix can be used in model (2.4). For example, the entry $d_{ijk}$ could be set to 1 if student $k$ takes at least one class with teacher $j$ and 0 otherwise. With this formulation, if a student has the teacher for multiple classes, the benefit of the same teacher to that student only occurs once.

The fixed effect vector, including the parameters at both student and teacher levels, is $(\boldsymbol{\gamma}', \boldsymbol{\theta}')'$. The parameter of interest is $\theta_p$, corresponding to the treatment effect. For the model in (2.4),

$$\boldsymbol{\Sigma}_i = \text{Cov}(\mathbf{Y}_i | \mathbf{X}_i, \mathbf{D}_i) = \sigma_s^2 \mathbf{J}_{n_i} + \sigma_t^2 \mathbf{D}_i \mathbf{D}_i' + \sigma_\eta^2 \mathbf{I}_{n_i}.$$

When $(\boldsymbol{\gamma}', \boldsymbol{\theta}')'$ is estimable, the generalized least squares estimator is

$$\begin{bmatrix} \hat{\boldsymbol{\gamma}} \\ \hat{\boldsymbol{\theta}} \end{bmatrix} = \left( \sum_{i=1}^{a} \begin{bmatrix} \mathbf{B}_i' \\ (\mathbf{D}_i \mathbf{X}_i)' \end{bmatrix} \boldsymbol{\Sigma}_i^{-1} \begin{bmatrix} \mathbf{B}_i & \vdots & \mathbf{D}_i \mathbf{X}_i \end{bmatrix} \right)^{-1} \sum_{i=1}^{a} \begin{bmatrix} \mathbf{B}_i' \\ (\mathbf{D}_i \mathbf{X}_i)' \end{bmatrix} \boldsymbol{\Sigma}_i^{-1} \mathbf{Y}_i.$$

The information matrix for a given design $\mathbf{X}$ and student assignment $\mathbf{D} = [\mathbf{D}_1' \cdots \mathbf{D}_a']'$ is

$$\mathcal{I}_S(\hat{\boldsymbol{\gamma}}, \hat{\boldsymbol{\theta}}, \mathbf{X}, \mathbf{D}) = \sum_{i=1}^{a} \begin{bmatrix} \mathbf{B}_i' \\ (\mathbf{D}_i \mathbf{X}_i)' \end{bmatrix} \boldsymbol{\Sigma}_i^{-1} \begin{bmatrix} \mathbf{B}_i & \vdots & \mathbf{D}_i \mathbf{X}_i \end{bmatrix}.$$

If there are no student-level covariates $\mathbf{B}_i$, the information simplifies to

$$(2.5) \qquad \mathcal{I}_S(\hat{\boldsymbol{\theta}}, \mathbf{X}, \mathbf{D}) = \sum_{i=1}^{a} \mathbf{X}_i' \mathbf{D}_i' \boldsymbol{\Sigma}_i^{-1} \mathbf{D}_i \mathbf{X}_i.$$

For the student model, we are primarily interested in the $(p, p)$ entry of the matrix $\text{Cov}(\hat{\boldsymbol{\theta}} | \mathbf{X}, \mathbf{D})$, the variance for our additive treatment effect, which we will approximate with $[\mathcal{I}_S(\hat{\boldsymbol{\theta}}, \mathbf{X}, \mathbf{D})]^{-1}$ (when the inverse exists). For specific values of $\mathbf{D}_i$, we can find the expected value of the information.

**3. Efficiencies of randomization designs.**  In this section we examine the variance of the treatment coefficient for teachers and students under three possible designs. One design is held to be more efficient than another if the variance of the treatment effect is smaller, given that the numbers of schools, teachers and students are the same in each design. To facilitate theoretical comparison of the designs, we make several simplifying assumptions. For each design, assume that each of the $a$ schools has the same number of teachers, $m_i = m$, where $m$ is even, and the same number of students, $n_i = n$. This scenario is a good approximation if we assume that schools in the study have been stratified by size, and models are fit separately to each stratum. Also assume that $\mathbf{x}'_{ij} = [1 \ \ x_{ij2}]$, where $x_{ij2}$ is the treatment indicator, and that there are no student covariates available. In practice, any important available covariates should be used to improve the precision of the design, and randomization is employed to remove residual biases [Cox and Reid (2000), page 33]. In Section 5 we present a computer program that can be used if the $m_i$'s and $n_i$'s are unequal.

In order to examine the variance of the treatment coefficient, we calculate the information matrix and the expected information for each design, for both the teacher-level and student-level responses. The inverse of the information matrix is the covariance matrix for the estimated fixed effects in each model, when those effects are estimable. We work with the information matrix rather than the covariance matrix because some of the designs can lead to a singular information matrix.

For any randomization that is done at the teacher or school level, the expected information depends on the $\mathbf{X}_i$ matrix. Let $\mathbf{X}_i = [\mathbf{1}_m \ \vdots \ \mathbf{R}_i]$, where the $j$th element of $\mathbf{R}_i$ is the treatment assignment of teacher $j$ from school $i$:

$$(3.1) \qquad \mathbf{R}_{ij} = \begin{cases} 1, & \text{if in experimental group,} \\ -1, & \text{if in control group.} \end{cases}$$

For the teacher model in (2.1), when randomization is employed, equations (2.2) and (2.3) show that the information is

$$(3.2) \qquad \mathcal{I}_T(\hat{\boldsymbol{\beta}}, \mathbf{X}) = \sum_{i=1}^{a} \left\{ \frac{1}{\sigma_\varepsilon^2} \begin{bmatrix} m & \mathbf{1}'\mathbf{R}_i \\ \mathbf{1}'\mathbf{R}_i & m \end{bmatrix} - \frac{\sigma_v^2}{\sigma_\varepsilon^2(\sigma_\varepsilon^2 + \sigma_v^2 m)} \begin{bmatrix} m^2 & m\mathbf{1}'\mathbf{R}_i \\ m\mathbf{1}'\mathbf{R}_i & \mathbf{1}'\mathbf{R}_i\mathbf{R}_i'\mathbf{1} \end{bmatrix} \right\}.$$

For the student model in (2.4), the information in (2.5) depends on how students are assigned to classes within each school. For a given assignment of students to teachers, the information is

$$\mathcal{I}_S(\hat{\boldsymbol{\theta}}, \mathbf{X}, \mathbf{D}) = \sum_{i=1}^{a} \begin{pmatrix} \mathbf{1}'\mathbf{D}_i'\boldsymbol{\Sigma}_i^{-1}\mathbf{D}_i\mathbf{1} & \mathbf{1}'\mathbf{D}_i'\boldsymbol{\Sigma}_i^{-1}\mathbf{D}_i\mathbf{R}_i \\ \mathbf{R}_i'\mathbf{D}_i'\boldsymbol{\Sigma}_i^{-1}\mathbf{D}_i\mathbf{1} & \mathbf{R}_i'\mathbf{D}_i'\boldsymbol{\Sigma}_i^{-1}\mathbf{D}_i\mathbf{R}_i \end{pmatrix}.$$

For each design, the expected information for $\theta_2$ for a given assignment of students to teachers is

$$\mathrm{E}[\mathcal{I}_S(\hat{\theta}_2, \mathbf{X}, \mathbf{D})|\mathbf{D}] = \sum_{i=1}^{a}\{\mathrm{tr}[\mathbf{D}_i'\mathbf{\Sigma}_i^{-1}\mathbf{D}_i\,\mathrm{Cov}(\mathbf{R}_i)] + \mathrm{E}[\mathbf{R}_i']\mathbf{D}_i'\mathbf{\Sigma}_i^{-1}\mathbf{D}_i\,\mathrm{E}[\mathbf{R}_i]\}.$$
(3.3)

3.1. *Design 1: randomize schools.* In the first design we randomize at the school level, assigning half of the available schools to each treatment. If school $i$ is randomized to the treatment group, then all teachers in the school $i$ are in the treatment group and $\mathbf{R}_i = \mathbf{1}_m$. Likewise, if school $i$ is randomized to the control group, then $\mathbf{R}_i = -\mathbf{1}_m$. Thus, under this design, $P(\mathbf{R}_i = \mathbf{1}_m) = P(\mathbf{R}_i = -\mathbf{1}_m) = 1/2$. Also, $\sum_{i=1}^{a} \mathbf{1}'\mathbf{R}_i = 0$, $\mathbf{R}_i\mathbf{R}_i' = \mathbf{J}_m$, $\mathrm{E}[\mathbf{R}_i] = 0$, and $\mathrm{Cov}(\mathbf{R}_i) = \mathbf{J}_m$.

For the teacher model, the information matrix from (2.2) and (3.2) for any realization of randomization with an equal number of schools in each treatment is

$$\mathcal{I}_{T1}(\hat{\boldsymbol{\beta}}, \mathbf{X}) = \frac{ma}{\sigma_\varepsilon^2 + \sigma_v^2 m}\mathbf{I}_2.$$

Therefore, the variance of the treatment variable for the teacher model when randomization is by school is $(\sigma_\varepsilon^2 + \sigma_v^2 m)/(ma)$.

For the student model, we have from equation (2.4) that the treatment effect is $\theta_2$ if there are no student covariates and $\boldsymbol{\theta} = (\theta_1, \theta_2)'$. When teachers are randomized by school, the information of the treatment indicator from (3.3) is

$$\mathcal{I}_{S1}(\hat{\theta}_2, \mathbf{X}, \mathbf{D}) = \sum_{i=1}^{a} \mathrm{tr}[\mathbf{D}_i'\mathbf{\Sigma}_i^{-1}\mathbf{D}_i\mathbf{J}_m]. \tag{3.4}$$

When all teachers in a school are randomized to the same treatment, the information does not depend on $\mathbf{X}$. Consequently, the expected information is attained for every realization of the design.

3.2. *Design 2: randomize teachers within schools.* In the second design we randomly assign half of the teachers at each school to the experimental treatment, and the other half to the control treatment. With $m$ teachers at each of the $a$ schools, each school will have the same form for the $\mathbf{X}_i$ matrix, with $x_{ij2} = 1$ for half of the teachers, and $x_{ij2} = -1$ for the other half of the teachers. For any vector $\mathbf{r}_i$ with entries 1 and $-1$ and with $\mathbf{1}'\mathbf{r}_i = 0$, we have $P(\mathbf{R}_i = \mathbf{r}_i) = \binom{m}{m/2}^{-1}$. Consequently, $\mathbf{1}'\mathbf{R}_i = 0$, $\mathrm{E}[\mathbf{R}_i] = 0$ and $\mathrm{Cov}(\mathbf{R}_i) = (m\mathbf{I}_m - \mathbf{J}_m)/(m-1)$.

For the teacher model, the information matrix from (3.2) for any realization of this randomization is

$$\mathcal{I}_{T2}(\hat{\boldsymbol{\beta}}, \mathbf{X}) = ma \begin{pmatrix} \dfrac{1}{(\sigma_\varepsilon^2 + \sigma_v^2 m)} & 0 \\ 0 & \dfrac{1}{\sigma_\varepsilon^2} \end{pmatrix}.$$

For the student model, the information for a specific randomization depends on the assignment of students to classes. The information can be zero in some cases, as will be shown in Section 3.5. The expected treatment information over all randomizations for a given assignment of students to classes is, from (3.3),

$$(3.5) \qquad \mathrm{E}[\mathcal{I}_{S2}(\hat{\theta}_2, \mathbf{X}, \mathbf{D})|\mathbf{D}] = \sum_{i=1}^{a} \frac{1}{m-1} \mathrm{tr}[\mathbf{D}_i' \boldsymbol{\Sigma}_i^{-1} \mathbf{D}_i (m\mathbf{I}_m - \mathbf{J}_m)].$$

3.3. *Design 3: completely randomized design of teachers.* In the third design we randomize half of the $ma$ teachers, regardless of school, to the experimental group, with the other half in the control group. In Sections 3.1 and 3.2 the variance of $\hat{\boldsymbol{\beta}}_p$ was the same for the teacher model regardless of which schools or teachers were randomly assigned to the treatment group. For the completely randomized design, school $i$ may have between 0 and $m$ teachers in the treatment group. Under this design, $\mathrm{E}[\mathbf{R}_i] = \mathbf{0}$ and $\mathrm{Cov}(\mathbf{R}_i) = (ma\mathbf{I}_m - \mathbf{J}_m)/(ma - 1)$.

For the teacher model, the expected information matrix from (3.2) for this design is

$$\mathrm{E}[\mathcal{I}_{T3}(\hat{\boldsymbol{\beta}}, \mathbf{X})] = \frac{ma}{\sigma_\varepsilon^2 + \sigma_v^2 m} \begin{bmatrix} 1 & 0 \\ 0 & 1 + \dfrac{(m-1)ma}{ma-1} \dfrac{\sigma_v^2}{\sigma_\varepsilon^2} \end{bmatrix}.$$

For the student model, the expected treatment information from (3.3) is

$$(3.6) \qquad \mathrm{E}[\mathcal{I}_{S3}(\hat{\theta}_2, \mathbf{X}, \mathbf{D})|\mathbf{D}] = \frac{1}{ma-1} \sum_{i=1}^{a} \mathrm{tr}[\mathbf{D}_i' \boldsymbol{\Sigma}_i^{-1} \mathbf{D}_i (ma\mathbf{I}_m - \mathbf{J}_m)].$$

3.4. *Comparison of designs.* Table 1 gives the expected information for the treatment coefficient for the designs discussed in Sections 3.1–3.3. Since each expected information matrix is diagonal, the anticipated variance of $\hat{\beta}_2$ or $\hat{\theta}_2$ will be the reciprocal of the information (when it is nonzero).

For teacher-level assessments, as shown in Moerbeek, van Breukelen and Berger (2000), the expected information is highest when randomization is done within the schools. The information is lowest for design 1, when teachers are randomized by school. Depending on the realization of randomization, the efficiency of the completely randomized design will be between that of randomizing by teacher within schools and randomizing by school.

For student-level assessments, the expected information of $\hat{\theta}_2$ given $\mathbf{D}_i$ is of the form

$$\sum_{i=1}^{a} \text{tr}[\mathbf{D}_i' \mathbf{\Sigma}_i^{-1} \mathbf{D}_i (k_1 \mathbf{I}_m + k_2 \mathbf{J}_m)],$$

for $k_1$, $k_2$ specified in Table 1. In general, $\mathbf{D}_i$ depends on the assignment of students to classes in each school, and a design that is most efficient for teacher-level assessments may not be efficient for student-level assessments. Because $\mathbf{\Sigma}_i$ depends on $\mathbf{D}_i$, in general, the expressions for expected treatment information in the student model need to be computed numerically. Simplifications are possible for some special cases of the student model when $\mathbf{D}_i$ has specified characteristics, and we examine one of these in the next section.

Other criteria and models may be used when evaluating designs. Since we are interested in measuring program impact on both teachers and students, we may choose to think of the information as a weighted average of the information from the two models: $\alpha \mathcal{I}_T(\hat{\beta}_2, \mathbf{X}) + (1-\alpha)\mathcal{I}_S(\hat{\theta}_2, \mathbf{X}, \mathbf{D})$, where $\alpha$ is chosen to reflect the relative importance of the two responses. The setup may also be extended to allow more hierarchical factors. For example, if schools are nested in districts, we could also consider randomizing the treatment by district. With even fewer clusters, this design would be less efficient for estimating teacher and student effects than design 1.

3.5. *Information for student model when $\mathbf{D}_i$ is balanced.* In this section we examine the special case when the assignment of students to classes is done in a balanced way. We define a balanced assignment to be one in which each student takes $c$ classes from different teachers, and each of the possible $\binom{m}{c}$ assignments of $c$ teachers to a student occurs with the same number of students. Consequently, in a balanced design, each class has $nc/m$ students. We show in Supplement A [Jenney and Lohr (2008a)] that, for a balanced assignment $\mathbf{D}_i$,

$$\text{(3.7)} \qquad \text{tr}(\mathbf{D}_i'\mathbf{\Sigma}_i^{-1}\mathbf{D}_i\mathbf{J}_m) = \frac{mnc^2}{mn\sigma_s^2 + nc^2\sigma_t^2 + m\sigma_\eta^2},$$

and

$$\text{(3.8)} \qquad \begin{aligned} m\,\text{tr}(\mathbf{D}_i'\mathbf{\Sigma}_i^{-1}\mathbf{D}_i) &= \text{tr}(\mathbf{D}_i'\mathbf{\Sigma}_i^{-1}\mathbf{D}_i\mathbf{J}_m) \\ &\quad + \frac{m(m-1)nc(m-c)}{m(m-1)\sigma_\eta^2 + nc(m-c)\sigma_t^2}. \end{aligned}$$

TABLE 1
*Comparison of expected information for treatment variable*

| | **Expected treatment information** | |
|---|---|---|
| **Randomization** | **Teacher model** | **Student model** |
| By school | $\frac{ma}{\sigma_\varepsilon^2 + \sigma_v^2 m}$ | $\sum_{i=1}^{a} \text{tr}[\mathbf{D}_i' \mathbf{\Sigma}_i^{-1} \mathbf{D}_i \mathbf{J}_m]$ |
| Within school | $\frac{ma}{\sigma_\varepsilon^2}$ | $\frac{1}{m-1} \sum_{i=1}^{a} \text{tr}[\mathbf{D}_i' \mathbf{\Sigma}_i^{-1} \mathbf{D}_i (m\mathbf{I}_m - \mathbf{J}_m)]$ |
| CRD | $\frac{ma}{(\sigma_\varepsilon^2 + \sigma_v^2 m)}[1 + \frac{(m-1)ma}{ma-1} \frac{\sigma_v^2}{\sigma_\varepsilon^2}]$ | $\frac{1}{ma-1} \sum_{i=1}^{a} \text{tr}[\mathbf{D}_i' \mathbf{\Sigma}_i^{-1} \mathbf{D}_i (ma\mathbf{I}_m - \mathbf{J}_m)]$ |

From (3.4) and (3.7), then, for any realization of the randomize-by-school design,

$$\mathcal{I}_{S1}(\hat{\theta}_2, \mathbf{X}, \mathbf{D}) = \frac{ac^2 n}{n\sigma_s^2 + c^2\sigma_t^2 n/m + \sigma_\eta^2}.$$

From (3.5), (3.7) and (3.8) the expected information for the randomize within schools design, given $\mathbf{D}$, is

$$\text{E}[\mathcal{I}_{S2}(\hat{\theta}_2, \mathbf{X}, \mathbf{D})|\mathbf{D}] = \frac{amnc(m-c)}{nc(m-c)\sigma_t^2 + m(m-1)\sigma_\eta^2}.$$

From (3.6), (3.7) and (3.8) the expected information for the completely randomized design, given $\mathbf{D}$, is

$$\begin{aligned} \text{E}[\mathcal{I}_{S3}(\hat{\theta}_2, \mathbf{X}, \mathbf{D})|\mathbf{D}] &= \frac{1}{ma-1}\left(\frac{a(a-1)mnc^2}{mn\sigma_s^2 + nc^2\sigma_t^2 + m\sigma_\eta^2}\right) \\ &\quad + \frac{1}{ma-1}\left(\frac{a^2 m(m-1)nc(m-c)}{nc(m-c)\sigma_t^2 + m(m-1)\sigma_\eta^2}\right). \end{aligned}$$

3.5.1. *Information when each student takes one class from each teacher.* When $c = m$, that is, each student takes a class from each of the $m$ teachers, then

$$\mathcal{I}_{S1}(\hat{\theta}_2, \mathbf{X}, \mathbf{D}) = \frac{am^2 n}{\sigma_\eta^2 + n(\sigma_s^2 + m\sigma_t^2)},$$

$$\text{E}[\mathcal{I}_{S2}(\hat{\theta}_2, \mathbf{X}, \mathbf{D})|\mathbf{D}] = 0,$$

$$\text{E}[\mathcal{I}_{S3}(\hat{\theta}_2, \mathbf{X}, \mathbf{D})|\mathbf{D}] = \frac{a(a-1)m^2 n}{(ma-1)(n\sigma_s^2 + nm\sigma_t^2 + \sigma_\eta^2)}.$$

Thus, in the design in which teachers are randomized within schools to the treatment, we cannot even estimate the desired treatment effect at the student level. In design 3 the expected information is positive, but it is possible to have a randomization in which $\theta_2$ is not estimable. The expected information from design 3 is less than that from design 1. When $c = m$, the most efficient design for estimating $\theta_2$ is design 1, in which all the teachers in a school are randomized to the same treatment. In this situation, then, design 1 is the most efficient design for estimating the effect of the intervention on the students. When examining the effect on teachers, we found that design 1 was the least efficient design.

3.5.2. *Information when each student takes one class.* For the special case in which each student takes a class with one teacher and each class has the same number, $n/m$, of students, then the information under any realization of design 1 and the expected information under designs 2 and 3 are

$$\mathcal{I}_{S1}(\hat{\theta}_2, \mathbf{X}, \mathbf{D}) = \frac{an}{\sigma_\eta^2 + \sigma_t^2 n/m + n\sigma_s^2},$$

$$\mathrm{E}[\mathcal{I}_{S2}(\hat{\theta}_2, \mathbf{X}, \mathbf{D})|\mathbf{D}] = \frac{an}{\sigma_t^2 n/m + \sigma_\eta^2},$$

$$\mathrm{E}[\mathcal{I}_{S3}(\hat{\theta}_2, \mathbf{X}, \mathbf{D})|\mathbf{D}] = \frac{1}{ma-1}\left(\frac{a(a-1)n}{n\sigma_s^2 + \sigma_t^2 n/m + \sigma_\eta^2}\right)$$
$$+ \frac{1}{ma-1}\left(\frac{a^2 n(m-1)}{\sigma_t^2 n/m + \sigma_\eta^2}\right).$$

In this case, design 2, when teachers are randomized within schools to the treatment, is the most efficient for estimating the treatment effect at the student level. Design 2 is also most efficient when examining the treatment effect at the teacher level.

3.5.3. *Information for student model in other situations.* When $c$ is between 2 and $m-1$, the relative efficiency of design 1 to design 2 depends on the values of the variance components $\sigma_\eta^2$ and $\sigma_s^2$. A necessary and sufficient condition for

$$\mathrm{E}[\mathcal{I}_{S2}(\hat{\theta}_2)] \geq \mathcal{I}_{S1}(\hat{\theta}_2)$$

is that

$$n(m-c)\sigma_s^2 \geq m(c-1)\sigma_\eta^2.$$

In practice, we expect $n$ to be large relative to $m$ and $\sigma_s^2 > 0$, so in most educational situations we would expect to have higher information for the estimated treatment effect on students from design 2 than from design 1. Note, though, that if $c$ is even, it is possible for a particular realization of design 2 to have information 0.

**4. Contamination, noncompliance and attrition.** Often, because of the nature of an intervention, a cluster-randomized trial is preferable to a completely randomized design or randomization within clusters. Contamination may be a concern in any experiment where subjects are clustered. For example, in medical informatics studies on decision support systems, a clinician treating both control and intervention patients may gain knowledge from the system when treating intervention patients and apply that knowledge to the treatment of the control patients [Chuang, Hripcsak and Heitjan (2002)]. Social science interventions are often community-based, and also susceptible to control group contamination when the control and treatment groups are mixed within clusters. In an educational setting, it may be difficult to limit the contamination of control groups when the treatment is applied at the teacher level and teachers are randomized to the treatment within schools. A discussion among teachers about teaching methods or lesson plans could influence a control group teacher to adopt elements of the treatment when teaching. Control group contamination can decrease the measurable effect of the treatment. If control group contamination is expected to be moderate to large, a cluster randomized trial is recommended [Moerbeek (2005)].

Likewise, noncompliance to the treatment within the experimental group and attrition can also reduce power to detect treatment effects. An experimental group teacher can be noncompliant by following the control group treatment, resulting in more similarity between control and experimental group. In a multi-semester evaluation, teachers or students may change schools or drop out of the study for other reasons, and this attrition may affect the power of the original design. In a study of noncompliance, Jo (2002) notes that higher power to detect treatment effects can also be obtained for within-cluster randomization (design 2) when fewer subjects are assigned to the intervention (unbalanced design).

Moerbeek (2005) studies effects of contamination on power in an "intent-to-treat" analysis for the teacher-level model. For that model, she argues that, for design 2, when $100q\%$ ($0 \leq q \leq 1$) of the control teachers follow the experimental treatment instead, the effect on power is equivalent to multiplying the variance of the estimated treatment coefficient by $(1-q)^{-2}$.

In this section we explore effects of contamination when it is known which teachers in the control group have been contaminated and explore effects of multicollinearity on the expected information from the designs studied in Section 3. Similar methodology could be applied to the treatment group to model noncompliance. Suppose that the teacher model is as in Section 2.1, with the modification that $\mathbf{x}_{ij}$ is a $(p+1)$-vector of covariates for the teacher. The additional covariate, the contamination indicator of an individual teacher, can be described as a Bernoulli random variable, where probability of contamination depends on the ratio of treatment teachers to control group teachers at a particular school. We will outline the model where control group teachers are subject to contamination. Then the new indicator, $C_{ij}$, equals 1 if teacher $j$ from school $i$ is in the control group and is contaminated, and equals zero otherwise. Let $C_{ij} = (1 - R_{ij})Z_{ij}/2$, where $Z_{ij} \sim \text{Bin}(1, (\mathbf{1}'\mathbf{R}_i + m)q/m)$, and $R_{ij}$ is defined in (3.1). For design 2, $0 \leq q \leq 1$; for design 3, $0 \leq q \leq 1/2$. Then when $\text{E}[\mathbf{R}_i] = 0$,

$$\text{E}[\mathbf{C}_i] = \frac{q}{2m}(m\mathbf{1}_m - \text{Cov}(\mathbf{R}_i)\mathbf{1}_m),$$

which depends on the randomization design.

For the teacher model, we still examine the $p$th element of $\boldsymbol{\beta}$, $\boldsymbol{\beta}_{p-1}$, which is the parameter of interest for assessing the effect of the treatment, but the estimate of the treatment effect may be influenced by the new contamination covariate. In the simple model where there are no additional teacher covariates, $\mathbf{X}_i = [\mathbf{1}_m \; \vdots \; \mathbf{R}_i \; \vdots \; \mathbf{C}_i]$, where $\mathbf{C}_i$ is the column of contamination indicators.

For design 1, randomize-schools, $\mathbf{C}_i = \mathbf{0}$ for all $i$ so contamination has no effect. For design 2, $\text{E}[\mathbf{R}_i] = \mathbf{0}$, $\text{Cov}(\mathbf{R}_i) = (m\mathbf{I}_m - \mathbf{J}_m)/(m-1)$, and $\text{E}[\mathbf{C}_i] = (q/2)\mathbf{1}_m$. For a general matrix $\mathbf{G}_i$,

$$\text{E}[\mathbf{X}_i'\mathbf{G}_i\mathbf{X}_i|\mathbf{G}_i]$$
$$= \mathbf{E} = \begin{bmatrix} \text{tr}(\mathbf{G}_i\mathbf{J}_m) & 0 & \frac{q}{2}\text{tr}(\mathbf{G}_i\mathbf{J}_m) \\ 0 & \text{tr}[\mathbf{G}_i\text{Cov}(\mathbf{R}_i)] & -\frac{q}{2}\text{tr}[\mathbf{G}_i\text{Cov}(\mathbf{R}_i)] \\ \frac{q}{2}\text{tr}(\mathbf{G}_i\mathbf{J}_m) & -\frac{q}{2}\text{tr}[\mathbf{G}_i\text{Cov}(\mathbf{R}_i)] & \text{tr}[\mathbf{G}_i\{\frac{q^2}{4}[\mathbf{J}_m + \text{Cov}(\mathbf{R}_i)] + \frac{q(1-q)}{2}\mathbf{I}_m\}] \end{bmatrix}.$$

When the inverse exists, the $(2,2)$ entry of $\mathbf{E}^{-1}$ is

$$\frac{1}{\text{tr}[\mathbf{G}_i\text{Cov}(\mathbf{R}_i)]}\left\{1 + \frac{q}{2(1-q)}\frac{\text{tr}[\mathbf{G}_i\text{Cov}(\mathbf{R}_i)]}{\text{tr}(\mathbf{G}_i)}\right\}.$$

Taking $\mathbf{G}_i = \mathbf{V}_i^{-1}$, for the teacher model the anticipated variance per school in design 2 is inflated by

$$\left\{1 + \frac{q}{2(1-q)}\left(1 - \frac{\sigma_v^2}{\sigma_\varepsilon^2 + m\sigma_v^2}\right)^{-1}\right\}.$$

For the student model, taking $\mathbf{G}_i = \mathbf{D}_i\boldsymbol{\Sigma}_i^{-1}\mathbf{D}_i$ for a balanced $\mathbf{D}_i$ as described in Section 3.5, and using equations (3.7) and (3.8), we have that the anticipated variance per school is inflated by

$$\left\{1 + \frac{q}{2(1-q)}\frac{mn\sigma_s^2 + nc^2\sigma_t^2 + m\sigma_\eta^2}{(m-1)n\sigma_s^2 + nc^2\sigma_t^2 + (m-1)/(m-c)\sigma_\eta^2}\right\}.$$

For design 2, in each of the teacher and student models, the anticipated variance increases due to the contamination. The anticipated variance inflation for design 3 can be calculated similarly.

EXPERIMENTAL DESIGNS FOR MULTIPLE-LEVEL RESPONSES     15

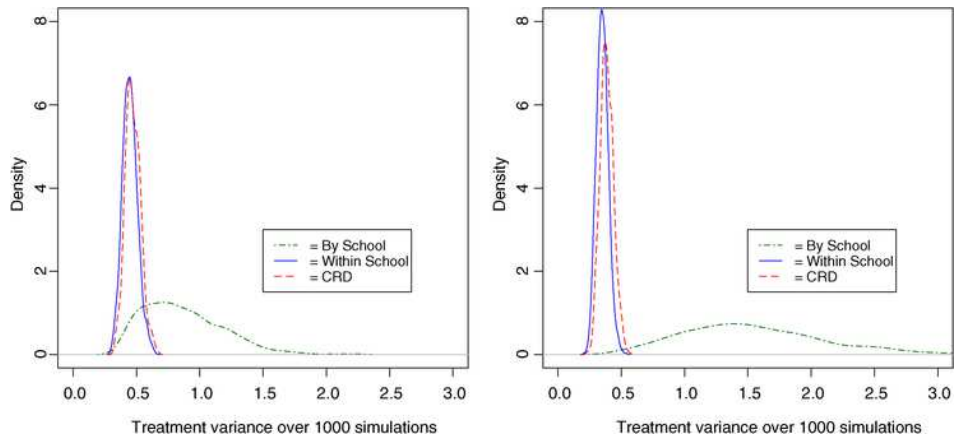

FIG. 1. *Teacher model: compare smoothed density of treatment variance for the three randomization designs. Left panel: $\sigma_v^2 = 1.6$, $\sigma_\varepsilon^2 = 14.4$. Right panel: $\sigma_v^2 = 4.8$, $\sigma_\varepsilon^2 = 11.2$. The randomize-schools design is less efficient when the between-school variance is a higher proportion of total variance (right panel).*

**5. Examining the distribution of the anticipated sample variance.** In Sections 3 and 4 we derived expected information matrices for the teacher and student models. For the student model, the information matrix depends in a complex way on the assignment of students to classes, and is analytically tractable only for special cases. In addition, it is possible for a specific realization of a design to have information that is far from its expected value. In this section we examine the distribution of the anticipated variance of the treatment effect computationally under various scenarios.

The macro *multleveldesign*, which is written for use with SAS software [SAS Institute Inc. (2008)] and is available in the supplementary material file posted at the journal website [Jenney and Lohr (2008b)], estimates the expected value and distribution of the anticipated variance of the treatment effect for inputted values of $a$, $n_1, \ldots, n_a$, $m_1, \ldots, m_a$ and the variance components. Unlike other programs such as Optimal Design [Liu et al. (2006); Raudenbush and Liu (2000)], *multleveldesign* handles multiple response levels and data that are not completely nested. The macro displays the distribution of the sample variance of the treatment effect for simulated data and gives the empirical power estimate for each randomization design of the student and teacher model.

We illustrate the macro with settings based on pilot data from Project Pathways. We plotted the simulated distribution of anticipated variance for the treatment variable at teacher and student levels using anticipated numbers of teachers and students in schools that would be available for the study. Each simulation in Figure 1 and Figure 2 generates data for 16 schools, with 8 teachers and 200 students at each school, according to the assumptions for the models in equations (2.1) and (2.4), with variance components stated in the figure captions. We also assume that each student takes two classes from teachers at their school, in order to generate the **D** matrix. For these examples, we randomly selected $c = 2$ teachers with replacement for each student. A study of efficiency of the models was conducted which set the intra-class correlation for the teacher model, $\rho = \sigma_v^2/(\sigma_v^2 + \sigma_\varepsilon^2)$, to be 0.1 or 0.3, and varied the other inputs of number of schools, teachers per school and students per school. Figure 3 and Figure 4 display results with model contamination in both the teacher and student model, with $q = 0.5$ in the contamination coefficient. Note that in Figure 4, the contamination can make designs 2 and 3 less efficient overall than design 1.

Given the possibility of control group contamination in Project Pathways, and based on the explorations in Figures 1–4 and other simulations under the assumptions given above, we believe a randomize-by-school design would be best. Under this scenario, with the estimated variance components given in the left panels of Figures 3 and 4, design 1 yields empirical standard deviation estimates of 0.9 for the teacher model, and 0.8 for the student model.

Although we did not include teacher or student covariates in the macro, they would likely be available to most researchers conducting an educational study and could be incorporated by taking $\sigma_v^2$ and $\sigma_\varepsilon^2$ to be variance components of residuals after adjusting for known covariates. This has the effect of reducing the cluster effects, since some of the school-to-school and teacher-to-teacher variability can be explained by covariates such as socio-economic status, years of teacher experience and other variables. The inclusion of covariates in the model should increase the power for detecting treatment differences [Bloom, Richburg-Hayes and Black (2005)].

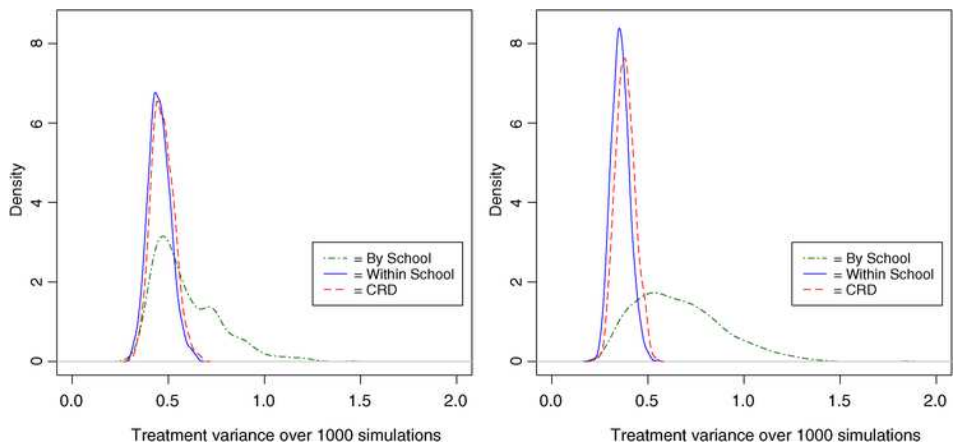

FIG. 2. *Student model: compare smoothed density of treatment variance for the three randomization designs. Left panel: $\sigma_s^2 = 1.6$, $\sigma_t^2 = 14.4$, $\sigma_\eta^2 = 14.4$. Right panel: $\sigma_s^2 = 4.8$, $\sigma_t^2 = 11.2$, $\sigma_\eta^2 = 11.2$. The randomize-schools design is nearly as efficient as the other two designs for the student model when the between-school variance is expected to be low (left panel).*

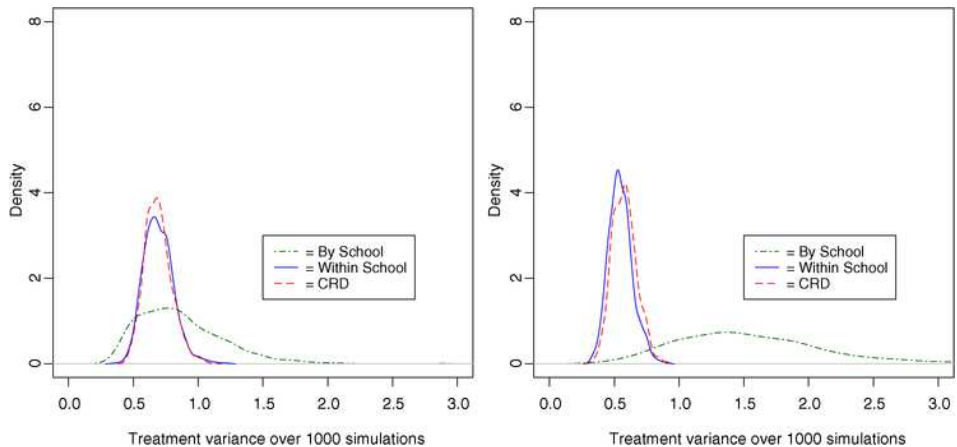

Fig. 3. *Contaminated teacher model: compare smoothed density of treatment variance for the three randomization designs, with $q = 0.5$. Left panel: $\sigma_v^2 = 1.6$, $\sigma_\varepsilon^2 = 14.4$. Right panel: $\sigma_v^2 = 4.8$, $\sigma_\varepsilon^2 = 11.2$. Contamination makes the randomize-within-schools and completely-randomized designs less efficient, but does not affect the randomize-schools design (as compared to Figure 1).*

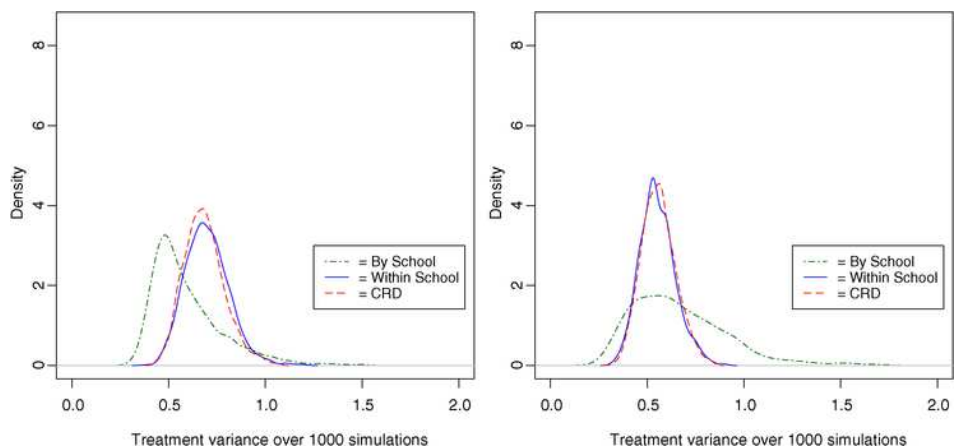

Fig. 4. *Contaminated student model: compare smoothed density of treatment variance for the three randomization designs, with $q = 0.5$. Left panel: $\sigma_s^2 = 1.6$, $\sigma_t^2 = 14.4$, $\sigma_\eta^2 = 14.4$. Right panel: $\sigma_s^2 = 4.8$, $\sigma_t^2 = 11.2$, $\sigma_\eta^2 = 11.2$. Contamination makes the randomize-within-schools and completely-randomized designs less efficient, but does not affect the randomize-schools design (as compared to Figure 2).*

**6. Conclusion.** The teacher-student conduit is crucial to the success of educational interventions such as Project Pathways, making it particularly important to evaluate the effect of the intervention on both teachers and students. There are clearly issues other than efficiency involved in the choice of evaluation design for Project Pathways. Among these are ease of implementation, school and teacher compliance, attrition and contamination, and the perceived fairness of the mechanism of randomization itself. Because many interventions such as Project Pathways encourage community-based activities, there can be a risk of contamination if the randomization is performed within schools or if a completely randomized design is used. These concerns, together with the need to measure impact on students, suggest that it may be beneficial for such studies to consider randomization at the school level even though that design may be less efficient for measuring the effect of the intervention on teachers.

In the simulations in Section 5 we assumed that students are assigned randomly to teachers within a school. In many design implementations, including Project Pathways, we would expect this assumption to be reasonable, since there is little reason to believe that choice of teacher would be influenced by the teacher's assignment to group. In other studies, however, self-selection of teachers by students may be more of a concern. If good students disproportionately choose teachers in the experimental group, the estimates of treatment effect will be biased. In such a situation we recommend a cluster-randomized trial, such as design 1.

We note that while this study used random effects models to describe dependence among teachers and students, if information about teacher interactions is available, dependence and contamination could alternatively be modeled using an adjacency matrix as in social network analysis [Wasserman and Faust (1994)]. Hoff (2003) presents random effects models to express the dependence found in social network data.

If the interest is mostly in the effect of the intervention on teachers, the most efficient designs would be to randomize assignment of treatments within schools. But these designs are not optimal for estimating the effect of the intervention on students—indeed, in certain cases when teachers are randomized within schools, the effect of the intervention on students is not even estimable. To estimate effects on students, it may be better to use a design in which randomization is performed at the school level when a student takes classes from multiple teachers at the school.

In this paper we discussed design issues in the context of an educational study. The results, however, are general and can be applied to any setting in which data are collected at multiple levels. One application, for example, would be clinical trials in which patients are treated by several health care practitioners.

**Acknowledgments.** The authors are grateful to the editor, associate editor and referee for their helpful comments and suggestions.

## SUPPLEMENTARY MATERIAL

**Supplement A: Proof of information for student model when $D_i$ is balanced** (DOI: 10.1214/08-AOAS216SUPPA; .pdf). Proofs of equations (3.7) and (3.8) appear in a supplementary file posted at the journal website.

**Supplement B: SAS program for simulation of anticipated variance** (DOI: 10.1214/08-AOAS216SUPPB; .zip). A SAS macro lets the user input simulation scenarios, including variance components, number of schools, number of teachers and students at each school, and contamination coefficient, then generates comparison graphs of the density of the anticipated variance for the three randomization designs under the teacher and student models, along with empirical power estimates.

## REFERENCES

BERK, R. A., LADD, H., GRAZIANO, H. and BAEK, J.-H. (2003). A randomized experiment testing inmate classification systems. *Criminology & Public Policy* **2** 215–242.
CRESMET, ARIZONA STATE UNIVERSITY (2007). Project Pathways (MSP). Available at http://cresmet.asu.edu/msp/newindex.shtml.
BLOOM, H., BOS, J. M. and LEE, S.-W. (1999). Using cluster random assignment to measure program impacts: Statistical implications for the evaluation of education programs. *Evaluation Review* **23** 445–469.
BLOOM, H., RICHBURG-HAYES, L. and BLACK, A. R. (2005). Using covariates to improve precision: Empirical guidance for studies that randomize schools to measure the impacts of educational interventions. MDRC Working Papers on Research Methodology. Available at http://www.mdrc.org/publications/417/abstract.html.
BORMAN, G. D., SLAVIN, R. E., CHEUNG, A., CHAMBERLAIN, A. M., MADDEN, N. A. and CHAMBERS, B. (2005). Success for all: First-year results from the national randomized field trial. *Educational Evaluation and Policy Analysis* **27** 1–22.
BORUCH, R. (2002). The virtues of randomness. *Education Next Fall* 37–41.
CHUANG, J.-H., HRIPCSAK, G. and HEITJAN, D. (2002). Design and analysis of controlled trials in naturally clustered environments: Implications for medical informatics. *Journal of the American Medical Informatics Association* **9** 230–238.
COOK, T. D. (2003). Why have educational evaluators chosen not to do randomized experiments? *Annals of the American Academy of Political and Social Science* **589** 114–149.
COOK, T. D. (2005). Emergent principles for the design, implementation, and analysis of cluster-based experiments in social science. *Annals of the American Academy of Political and Social Science* **599** 176–198.
COOK, T. D. and PAYNE, M. R. (2002). Objecting to the objections to using random assignment in educational research. In *Evidence Matters: Randomized Trials in Education Research* (F. Mosteller and R. Boruch, eds.) 150–178. Brookings Institution Press, Washington, DC.
COX, D. and REID, N. (2000). *The Theory of the Design of Experiments.* Chapman and Hall/CRC Press, Boca Raton, FL.

Demidenko, E. (2004). *Mixed Models*. Wiley, Hoboken, NJ. MR2077875

Gail, M. H., Mark, S. D., Carroll, R. J., Green, S. B. and Pee, D. (1996). On design considerations and randomization-based inference for community intervention trials. *Stat. Med.* **15** 1069–1092.

Gueron, J. (2005). Throwing good money after bad: A common error misleads foundations and policymakers. *Stanford Social Innovation Review Fall* 69–71.

Hoff, P. D. (2003). Random effects models for network data. Dynamic Social Network Modeling and Analysis: Workshop Summary and Papers. Available at http://www.nap.edu/openbook.php?record_id=10735&page=303.

Jenney, B. and Lohr, S. (2008a). Supplement to "Experimental designs for multiple-level responses, with application to a large-scale educational intervention." DOI: 10.1214/08-AOAS216SUPPA.

Jenney, B. and Lohr, S. (2008b). Supplement to "Experimental designs for multiple-level responses, with application to a large-scale educational intervention." DOI: 10.1214/08-AOAS216SUPPB.

Jo, B. (2002). Statistical power in randomized intervention studies with noncompliance. *Psychological Methods* **7** 178–193.

Johnson, T. (1998). Clinical trials in psychiatry: Background and statistical perspective. *Stat. Methods Med. Res.* **7** 209–234.

Liu, X., Spybrook, J., Congdon, R. and Raudenbush, S. (2006). Optimal design software for multi-level and longitudinal research v1.77. Available at http://sitemaker.umich.edu/group-based/optimal_design_software.

McCaffrey, D. F., Koretz, D., Louis, T. A. and Hamilton, L. (2004). Models for value-added modeling of teacher effects. *Journal of Educational and Behavioral Statistics* **29** 67–101.

Moerbeek, M. (2005). Randomization of clusters versus randomization of persons within clusters: Which is preferable? *Amer. Statist.* **59** 72–78. MR2113199

Moerbeek, M., van Breukelen, G. J. P. and Berger, M. P. F. (2000). Design issues for experiments in multilevel populations. *Journal of Educational and Behavioral Statistics* **25** 271–284.

Raudenbush, S. (1997). Statistical analysis and optimal design for cluster randomized trials. *Psychological Methods* **2** 173–185.

Raudenbush, S. and Liu, X. (2000). Statistical power and optimal design for multisite randomized trials. *Psychological Methods* **5** 199–213.

SAS Institute Inc. (2008). SAS/STAT 9.2 user's guide. SAS Institute Inc., Cary, NC.

Wasserman, S. and Faust, K. (1994). *Social Network Analysis: Methods and Applications*. Cambridge Univ. Press, Cambridge.

What Works Clearinghouse (2006). Evidence standards for reviewing studies. Available at http://ies.ed.gov/ncee/wwc/pdf/study_standards_final.pdf.

Department of Mathematics
and Statistics
Arizona State University
Tempe, Arizona 85287-1804
USA
E-mail: brenda.jenney@asu.edu
sharon.lohr@asu.edu